# An Automated System for Checking Lithography Friendliness of Standard Cells

I-Lun Tseng, *Senior Member*, *IEEE*, Yongfu Li, *Senior Member*, *IEEE*, Valerio Perez, Vikas Tripathi, Zhao Chuan Lee, and Jonathan Yoong Seang Ong

GLOBALFOUNDRIES Singapore Pte. Ltd.

*Abstract*—At advanced process nodes, lithography weakpoints can exist in physical layouts of integrated circuit designs even if the layouts pass design rule checking (DRC). Existence of lithography weakpoints in a physical layout can cause manufacturability issues, which in turn can result in yield losses. In our experiments, we have found that specific standard cells have tendencies to create lithography weakpoints after their cell instances are placed and routed, even though each of these cells does not contain any lithography weakpoint before performing placement and routing. In addition, our experiments have shown that abutted standard cell instances can induce lithography weakpoints. Therefore, in this paper, we propose methodologies that are used in a novel software system for checking standard cells in terms of the aforementioned lithography issues. Specifically, the software system is capable of detecting and sorting problematic standard cells which are prone to generate lithography weakpoints, as well as reporting standard cells that should not be abutted. Methodologies proposed in this paper allow us to reduce or even prevent the generation of undesirable lithography weakpoints during the physical synthesis phase of designing a digital integrated circuit.

## I. Introduction

A lithography weakpoint pattern is a geometric layout pattern which contains one or more polygon shapes, and the pattern is difficult to be produced correctly on silicon wafers during a semiconductor manufacturing process. At advanced process nodes (e.g., 22nm, 14nm, and 7nm nodes), lithography weakpoints can act as major factors to cause yield losses, especially under aggressive design rules [1]. Specifically, a routed layout design can contain lithography weakpoints even if it passes design rule checking (DRC). To efficiently detect lithography weakpoints existing in physical layouts, pattern matching based methodologies [2]–[4] and machine learning based techniques [5]–[6] have been proposed. Note that, for the reason of brevity, a lithography weakpoint is also referred to as a weakpoint in this paper.

Although commercial pattern matching and in-design weakpoint fixing tools have been developed and used in the semiconductor industry, there are weakpoints that cannot be fixed by these tools after the detailed routing stage is completed. Furthermore, for weakpoints that can be fixed, the weakpoint fixing stage can involve processes of standard cell shifting, standard cell flipping, ripup-and-rerouting [7], and/or surgical fixing [8]; these fixing processes can affect timing and power of circuit designs. Therefore, in the design of a digital integrated circuit, it is desirable to prevent the generation of unwanted lithography weakpoints during the physical synthesis phase.

From our physical design experiments, we have found that specific standard cells have tendencies to create weakpoints after their cell instances are placed and routed. For instance, in physical designs synthesized by using a commercial APR (automatic placement and routing) tool, our experiments have shown that improperly designed standard cells can cause synthesized layouts to contain many weakpoints. Specifically, in many routed designs, we have found a number of such problematic standard cells, and as high as 95% of a problematic cell's instances generated lithography weakpoints. Therefore, it is important to detect standard cells that are prone to generate weakpoints. Since methodologies proposed in this paper allow us to detect these problematic cells, circuit designers can either modify or redesign the problematic cells in order to reduce or prevent the generation of weakpoints during the physical synthesis phase of designing a digital integrated circuit.

In this paper, lithography friendliness of standard cells is considered in terms of (1) routing for the standard cells, and (2) horizontal abutments of standard cell instances. We define notations $PL_{i,j,k}$ and $LFR_{i,j,k}$ below before the problem description is presented.

*Definition*: For standard cell $i$, $PL_{i,j,k}$ is defined as the percentage of the cell's instances which induce type $j$ lithography weakpoints in a routed layout design containing $n$ instances of cell $i$; the routed layout design is denoted by design $k$. □

*Definition* (Lithography Friendliness of a Standard Cell in Terms of Routing): $LFR_{i,j,k}$ for standard cell $i$ with regard to type $j$ lithography weakpoints in routed design $k$ is defined as:

$$LFR_{i,j,k} = 100\% - PL_{i,j,k} \qquad □$$

For example, for a standalone standard cell which already contains a weakpoint before the routing stage is performed, the lithography friendliness of the cell in terms of routing can be either 0% or very close to 0%.

The goal of this research focuses on solving the following problem.

*Problem Description*: Given a standard cell library containing a number of standard cells, (1) detect problematic standard cells that are prone to generate lithography weakpoints if instances of these cells are placed and routed, and (2) report $LFR_{i,j,k}$ values for these problematic cells.

Since accurate values of $PL_{i,j,k}$ and $LFR_{i,j,k}$ can be difficult to obtain from a routed design, we use $PL'_{i,j,k}$ and $LFR'_{i,j,k}$, respectively, to represent their approximate values. Notations $PL'_{i,j,k}$ and $LFR'_{i,j,k}$ are defined as follows.

*Definition*: For standard cell $i$ in design $k$, $PL'_{i,j,k}$ is defined as:

$PL'_{i,j,k}$ = ( (the total number of the cell's instances whose PR boundary regions have partial overlaps with type $j$ lithography weakpoints) / (the total number of the cell's instances in design $k$) ) × 100% . □

*Definition*: $LFR'_{i,j,k}$ for standard cell $i$ with regard to type $j$ lithography weakpoints in routed design $k$ is defined as:

$$LFR'_{i,j,k} = 100\% - PL'_{i,j,k} \qquad \square$$

Consequently, a value of $LFR'_{i,j,k}$ is more pessimistic than its corresponding value of $LFR_{i,j,k}$ in a design. For instance, if the value of $LFR'_{i,j,k}$ equals 85%, the value of its corresponding $LFR_{i,j,k}$ must be greater than or equal to 85%. That is because some weakpoints can be induced by routed wires alone, but they are located above standard cell instances.

Note that, after those problematic cells have been detected, we can either modify their layouts (e.g., by adding routing blockages) or redesign those cells in order to prevent the generation of weakpoints during the physical synthesis phase. In other words, detecting those problematic cells is an essential step toward the prevention of lithography weakpoints. This paper proposes methodologies for solving the aforementioned problem, and these methodologies have been implemented in a software system. Moreover, based on computed $LFR'_{i,j,k}$ values for standard cells, problematic cells which should be modified or re-designed can be prioritized.

In addition to solving the aforementioned problem, the implemented software system has other features. In particular, the software system can be run in either GUI (Graphical User Interface) mode or batch mode. It also offers parallel execution of tasks which are selected by the user. Furthermore, the system is able to generate standard cell verification reports in a web-based format. The software system can be used for the verification of standard cells in terms of their lithography friendliness.

The remainder of this paper is organized as follows. Section II presents three major situations when undesirable lithography weakpoints can be created during the physical synthesis phase. Methodologies used in the automated system for checking lithography friendliness of standard cells are proposed in Section III. Experimental results are presented in Section IV. Finally, conclusions are drawn in Section V.

## II. PRELIMINARY

This section describes three major situations when lithography weakpoints are formed during the placement and routing stage. Although weakpoints can also be formed in other situations, our experiments have indicated that these

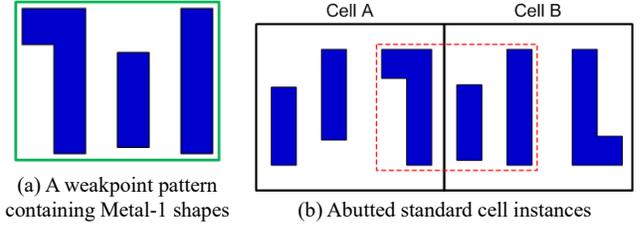

(a) A weakpoint pattern containing Metal-1 shapes  (b) Abutted standard cell instances

**Figure 1**. A weakpoint can be formed when two standard cell instances are abutted.

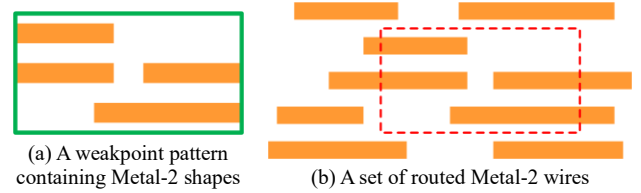

(a) A weakpoint pattern containing Metal-2 shapes  (b) A set of routed Metal-2 wires

**Figure 2**. A weakpoint can be formed by routed wires.

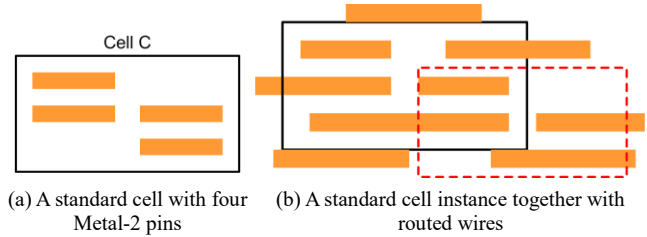

(a) A standard cell with four Metal-2 pins  (b) A standard cell instance together with routed wires

**Figure 3**. A weakpoint can be formed by layout shapes of a standard cell instance together with shapes created by a router.

three situations occurred frequently. Note that, in this section, we assume each standalone standard cell does not contain any weakpoint, although our software system is able to detect standalone cells which contain weakpoints before their instances are routed.

### A. Horizontally-Abutted Standard Cell Instances (HASC)

In a cell-based digital design flow, even though each standalone standard cell does not have any weakpoint, undesirable weakpoints can still be formed when standard cell instances are horizontally abutted. An example of a weakpoint pattern is shown in Fig. 1(a); the pattern contains three Metal-1 shapes, which are inside a pattern extent. As illustrated in Fig. 1(b), a weakpoint can be generated when two standard cell instances are abutted horizontally, even though there is no routed wire in the layout shown in Fig. 1(b) and each standalone cell does not contain any weakpoint. Note that the red dashed-outline rectangle in Fig. 1(b) denotes the marker of the weakpoint. For vertically abutted standard cell instances, on the other hand, they are usually not as common as horizontally-abutted cell instances to create weakpoints since the layout of a standard cell usually has a power rail at the top (bottom) and a ground rail at the bottom (top); both rails are usually drawn as wide metal lines. For the reason of brevity, the situation of horizontally abutted standard cell instances is named HASC.

*B. Routed Wires Alone*

In the second situation, undesirable weakpoints are induced by routed wires alone; these wires are usually created by a router. For instance, a weakpoint pattern containing a pattern extent and polygons on the Metal-2 layer is illustrated in Figure 2(a), and a group of routed Metal-2 wires shown in Figure 2(b) contains a weakpoint, which is marked by a red dashed-outline rectangle.

*C. Standard Cell Layout Shapes with Routed Shapes (SCRS)*

Lithography weakpoints can also be formed by layout shapes of a standard cell instance together with polygon shapes created by a router [9]; the layout shapes of the standard cell can be I/O pins of the cell, and the shapes created by the router can be routed wires and/or vias with metal enclosures. Fig. 3(a) shows an example of a standard cell which contains four Metal-2 pins. If we consider the weakpoint pattern shown in Fig. 2(a), the layout of the cell in Fig. 3(a) does not contain any weakpoint. However, after an instance of the cell is placed in a design and then routed, extra Metal-2 shapes can be added by the router as the example illustrated in Fig. 3(b); note that two I/O pins of the cell instance are connected with routed Metal-2 wires, and thus an unwanted weakpoint is generated. For the reason of brevity, the situation of standard cell layout shapes with routed shapes is named SCRS.

### III. SYSTEM OVERVIEW AND LITHOGRAPHY FRIENDLINESS CHECKING

An overview of our lithography friendliness checking system is illustrated in Fig. 4. The system takes a number of files as the input, and generates reports in a web-based format after verification tasks are completed. To perform all of verification tasks, the following input files are required:

- A standard cell library (e.g., in GDS and LEF file formats)
- A pattern matching rule deck for detecting lithography weakpoints [11]
- A router technology file
- A setting file for performing routing (e.g., settings for routing tracks) and for running the software system (e.g., the number of CPUs that the user intends to use)

To perform all of lithography friendliness verification tasks, the software system will generate three placement scenarios after it has been launched. These three placement scenarios are named (1) standalone-cell scenario, (2) auto-placed scenario, and (3) abutted-cell scenario; each of them are presented in the following subsections.

*A. Standalone-Cell Scenario*

The main purpose of generating a placement in the standalone-cell scenario is to detect weakpoint issues in the situation of SCRS; the situation has been described in Section II. Specifically, in the standalone-cell scenario, the placement contains instances of all standard cells which are obtained from a given standard cell library; each standard cell of the cell library has one and only one instance in the placement. Moreover, in the placement, each standard cell instance is located in one or more unique standard cell rows, and there is at least one empty row between neighboring cell instances. Furthermore, instances of multiple-height cells are also included in the placement. Fig. 5 illustrates an example of a placement in the standalone-cell scenario; the placement contains instances of three single-height cells and one double-height cell.

After the standalone-cell placement has been generated, the system performs partial pattern matching (PPM) [10] on the generated placement. However, the technique used in our system for generating partial patterns is different from the technique proposed in [10]. That is because, in our system, a solid polygon which is removed from the original weakpoint pattern is replaced by a "don't care" region. For the original weakpoint pattern shown in Fig. 2(a), Fig. 6 illustrates its four partial patterns generated by using our new PPM-based technique. Note that the orange dashed-outline rectangles in Fig. 6 represent "don't care" regions for the Metal-2 layer.

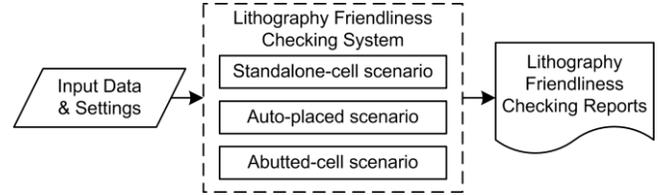

**Figure 4**. An overview of the lithography friendliness checking system.

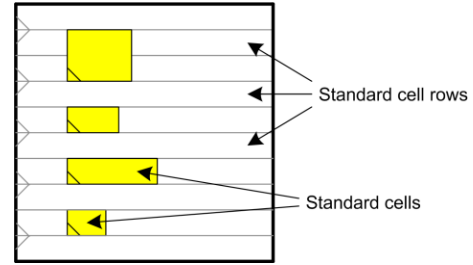

**Figure 5**. An example of a placement in the standalone cell scenario.

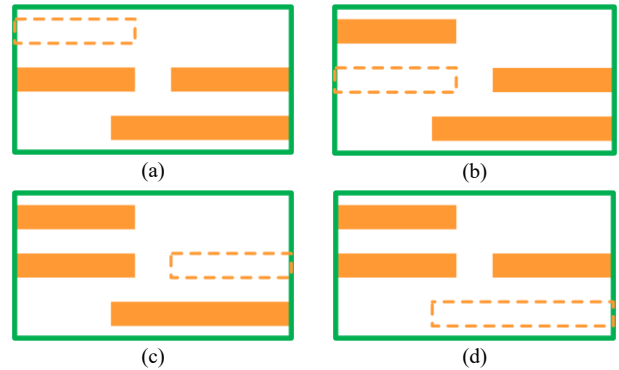

**Figure 6**. Four partial patterns generated from the original weakpoint pattern shown in Fig. 2(a) by using our new technique.

After performing pattern matching on a created standalone-cell placement by using generated partial patterns, PPM violation results can be obtained; these results indicate locations of standard cells which can induce weakpoints if instances of these standard cells are routed. For instance, the layout of Cell C shown in Fig. 3(a) has two PPM violations for the partial pattern illustrated in Fig. 6(d), indicating that instances of the cell can form weakpoints after they are routed. On the other hand, if the original PPM technique proposed in [10] had been used instead, the weakpoint issue existing in Cell C would not have been detected. Note that our new PPM-based technique is very efficient in detecting problematic cells in the situation of SCRS since no time-consuming routing stage is required.

### B. Auto-Placed Scenario

The major purpose of generating a placement in the auto-placed scenario is to compute $LFR'_{i,j,k}$ values for standard cell $i$ with regard to type $j$ lithography weakpoints. Standard cells which have low $LFR'_{i,j,k}$ values (e.g., ≤ 85%) should be modified or even redesigned since their cell instances had high chances of creating weakpoints after these instances are routed. By combining results generated from our software system for the standalone-cell and auto-placed scenarios, circuit designers will be able to know which problematic cells are the most likely to generate unwanted weakpoints, and they will also be able to identify locations relevant to these issues within the problematic cells.

Before generating a placement in the auto-placed scenario, our software system generates a netlist (e.g., in Verilog format) which contains all of standard cells from the given standard cell library, except specific types of cells (e.g., filler cells) that should not be included during the automatic placement stage. Furthermore, for each of the included cells, the netlist must contain an enough number (e.g., 800) of its instances. After the netlist has been generated, the software system will perform placement and routing based on a non-timing physical synthesis flow. After that, a lithography weakpoint detection process is executed on the routed physical layout according to the pattern matching rule deck (also known as the DRC Plus rule deck [11]) which is provided by the user. Finally, $LFR'_{i,j,k}$ values for standard cell $i$ with regard to type $j$ lithography weakpoints can then be computed.

### C. Abutted-Cell Scenario

The major goal of generating a placement in the abutted-cell scenario is to detect weakpoint issues in the situation of HASC; the situation has been described in Section II. Therefore, for this placement scenario, our software system generates a placement which contains all combinations of horizontally-abutted standard cell instances based on the given standard cell library. Our approach to the generation of abutted-cell placements is based on the technique proposed in [12]. However, our approach considers not only single-height cells, but also multiple-height cells. Furthermore, our approach takes placement constraints into account and only generates legal placements since specific standard cells may not be abutted.

After an abutted-cell placement has been generated, our system will perform lithography weakpoint detection according to the pattern matching (PM) rule deck provided by the user; note that it is not required to perform routing after the abutted-cell placement has been created. Next, if the lithography weakpoint detection result contains one or more PM violations, there must exist standard cell instances that cannot be abutted. Locations of PM violations thus indicate that relevant standard cell instances cannot be abutted. Designers can choose to either enforce new placement constraints on the cell instances, or redesign physical layouts of relevant standard cells.

## IV. EXPERIMENTAL RESULTS

The lithography friendliness verification software system was mainly developed in Python and Tcl languages. The system uses a commercial APR (automatic placement and routing) tool for performing placement and routing. It also uses a commercial PM (pattern matching) tool for performing lithography weakpoint detection and partial pattern matching. All of our experiments were carried out on a server which had 16 CPU cores.

Table I shows the information of three weakpoint patterns which were selected from an advanced process node of GLOBALFOUNDRIES; these three patterns ($j = 1-3$) were selected for the demonstration of our experiments since they were detected relatively frequently in our routed designs. Notice that the table shows the number of generated partial patterns for each original weakpoint pattern.

Table II shows the runtimes of our software system for checking an experimental standard cell library, which contained 832 standard cells. Since the software system is able to execute multiple tasks in parallel, the overall elapsed time for verifying the cell library is less than 4 hours. Additionally, Table III shows the results of checking the standard cell library by listing the information on top six standard cells which contributed the most numbers of lithography weakpoints. As shown in the table, the $LFR'_{i,1,1}$ value for Cell #1 ($i = 1$) in the design was 52.7%, indicating that around one in two of the cell's instances created type 1 lithography weakpoints during routing. Therefore, to reduce lithography weakpoints existing in a design, it is recommended to either modify or redesign the standard cell if the cell is frequently used in the design.

**TABLE I.** SELECTED WEAKPOINT PATTERNS FOR AN ADVANCED PROCESS NODE

| Type of Lithography Weakpoint Pattern ($j$) | Related Layers | Number of Partial Patterns |
|---|---|---|
| 1 | Via-1, Metal-2 | 5 |
| 2 | Via-1, Metal-2 | 4 |
| 3 | Metal-2, Via-2 | 4 |

**TABLE II.** RUNTIMES (ELAPSED TIMES) OF THE SOFTWARE SYSTEM FOR CHECKING A STANDARD CELL LIBRARY

|  | Standalone-Cell Scenario | Auto-placed Scenario | Abutted-Cell Scenario |
|---|---|---|---|
| Elapsed Time | 9 min. 5 sec. | 3 hrs 57 min. 16 sec. | 13 min. 4 sec. |

TABLE III. INFORMATION ON THE TOP SIX PROBLEMATIC STANDARD CELLS WHICH CONTRIBUTED THE MOST NUMBERS OF LITHOGRAPHY WEAKPOINTS IN A BENCHMARK DESIGN ($k = 1$, $n = 800$)

|  | Cell #1 | Cell #2 | Cell #3 | Cell #4 | Cell #5 | Cell #6 |
|---|---|---|---|---|---|---|
| Percentage of Total Instance Count | 0.125% | 0.125% | 0.125% | 0.125% | 0.125% | 0.125% |
| $LFR'_{i,1,1}$ | 52.7% | 57.0% | 57.3% | 58.9% | 99.3% | 99.8% |
| $LFR'_{i,2,1}$ | 100% | 100% | 100% | 100% | 80.2% | 80.5% |
| $LFR'_{i,3,1}$ | 99.8% | 99.8% | 99.5% | 99.8% | 97.8% | 98.3% |
| Minimum $LFR'_{i,j,k}$ | 52.7% | 57.0% | 57.3% | 58.9% | 80.2% | 80.5% |

TABLE IV. INFORMATION ON SIX MODIFIED STANDARD CELLS IN A BENCHMARK DESIGN ($k = 2$, $n = 800$)

|  | Cell #1 | Cell #2 | Cell #3 | Cell #4 | Cell #5 | Cell #6 |
|---|---|---|---|---|---|---|
| Percentage of Total Instance Count | 0.125% | 0.125% | 0.125% | 0.125% | 0.125% | 0.125% |
| $LFR'_{i,1,2}$ | 100% | 100% | 100% | 100% | 99.9% | 100% |
| $LFR'_{i,2,2}$ | 100% | 100% | 100% | 100% | 100% | 100% |
| $LFR'_{i,3,2}$ | 98.3% | 99.9% | 100% | 99.8% | 93.9% | 96.4% |
| Minimum $LFR'_{i,j,k}$ | 98.3% | 99.9% | 100% | 99.8% | 93.9% | 96.4% |

After modifying the six problematic cells listed in Table III, lithography weakpoints were reduced by 15% in a design, although the percentage of the combined instance count for the six cells in the design was only 0.75%. Modifications of physical layouts for these problematic cells were based on PPM results in the standalone-cell scenario since potential weakpoints had been located for each problematic cell during the process. Table IV shows the information on the six modified cells in the routed design. Note that the routed layout designs for Table III and Table IV were based on the same cell-level netlist and the same floorplan. As shown in Table IV, the lithography friendliness of the modified cells in terms of routing was improved.

## V. CONCLUSION

This paper proposes methodologies used in a novel software system for the detection of problematic standard cells which have tendencies to create lithography weakpoints during the physical synthesis phase. In addition, specific locations of potential weakpoints within standard cells can be identified. Furthermore, our experiments have shown that many lithography weakpoints can be prevented if physical layouts of those problematic cells are modified. Our future work includes the enhancement of the software system so that other checks, such as pin accessibility checks for standard cells, can be performed.